\documentclass[twocolumn]{aastex631}

\usepackage{amsmath}
\DeclareUnicodeCharacter{02BC}{'}
\usepackage{xcolor}
\begin{document}
\title{Volume Dependence of Helicity-based Eruption Diagnostics in Recurrent Coronal Jets}

\correspondingauthor{Jinhan Guo and Yang Guo}
\email{jinhan.guo@nju.edu.cn, guoyang@nju.edu.cn}

\author[0009-0009-1911-399X]{Chang Zhou}
\affiliation{School of Astronomy and Space Science and Key Laboratory of Modern Astronomy and
Astrophysics, Nanjing University, Nanjing 210023, People’s Republic of China}

\author[0000-0002-4205-5566]{Jinhan Guo}
\affiliation{School of Astronomy and Space Science and Key Laboratory of Modern Astronomy and
Astrophysics, Nanjing University, Nanjing 210023, People’s Republic of China}

\author[0000-0002-9293-8439]{Yang Guo}
\affiliation{School of Astronomy and Space Science and Key Laboratory of Modern Astronomy and
Astrophysics, Nanjing University, Nanjing 210023, People’s Republic of China}

\author[0000-0003-2569-1840]{Jiajia Liu}
\affiliation{National Key Laboratory of Deep Space Exploration/School of Earth and Space Sciences,
University of Science and Technology of China, Hefei 230026, China}

\author[0000-0001-7961-7617]{Zekun Lu}
\affiliation{Institute for Space-Earth Environmental Research, Nagoya University, Furocho,
Chikusa-ku, Nagoya, Aichi 464-8601, Japan}

\author[0000-0002-6641-8034]{Yuhang Gao}
\affiliation{School of Earth and Space Sciences, Peking University, Beijing 100871, People’s
Republic of China}

\begin{abstract}

Recurrent solar coronal jets originate from the same source region,
providing a useful setting for
studying the buildup, redistribution, and release of magnetic energy and helicity.
However, how these quantities can characterize magnetic evolution and the onset conditions of the eruptions remains unclear.
We perform a full three-dimensional (3D) thermodynamic magnetohydrodynamic simulation of recurrent
jets driven by converging motions imposed at the bottom boundary.
The simulation naturally produces two successive eruptions.
The current-carrying helicity $H_J$ decreases during the first jet and gradually recovers between
the two jets.
In contrast, the free magnetic energy $E_{\mathrm{free}}$ shows no clear accumulation after its
first release.
This difference indicates that $H_J$ provides information on magnetic evolution beyond that
provided by $E_{\mathrm{free}}$.
Furthermore, the commonly used helicity ratio $H_J/H_V$ reaches different peaks near the onset of
the two jets and is sensitive to the integration volume.
By comparison, the locally normalized quantity $H_J/\Phi_{\mathrm{closed}}^2$ reaches similar
peaks near the onset of both jets and is relatively insensitive to the choice among the tested
integration volumes, where
$\Phi_{\mathrm{closed}}$ is the single-polarity magnetic flux of the closed-field domain.
These results reveal the local nature of the eruptions and suggest that $H_J$ normalized by the
square of the closed magnetic flux may be a useful diagnostic of eruptive conditions.

\end{abstract}

\section{Introduction}\label{sec:1}

Solar coronal jets are ubiquitous transient phenomena characterized by rapid plasma ejections in the
solar atmosphere.
Most of them are driven by magnetic reconnection, which converts magnetic energy into plasma kinetic
and thermal energy (\citealp{Shibata1992_PASJ}; \citealp{Moore2010_ApJ}).
Studying coronal jets helps us understand several fundamental problems in solar physics, including
coronal heating \citep{Antolin2021}, solar wind generation \citep{Chitta2023_Science,Chitta2025_AA},
and the physical mechanisms of solar eruptions (\citealp{Sterling2015_Nature};
\citealp{Wyper2017_Nature}; \citealp{Gao2025_ApJL}).
In addition, observations suggest that coronal jets can trigger and contribute to larger-scale
eruptions such as coronal
mass ejections (\citealp{Liujiajia_2015ApJ}).
Coronal jets are also associated with the generation of magnetohydrodynamic (MHD) waves,
which may further contribute to coronal heating and solar-wind acceleration
(\citealp{Lynch2014_SoPh}; \citealp{Gao2026_NSR}).
The connection between magnetic reconnection and wave generation in coronal jets has recently been
investigated through numerical simulations (e.g., \citealp{He2021_ApJL};
\citealp{YangLiping_2025ApJL}).

Recurrent coronal jets occur repeatedly at the same source site and have been observed in active
regions \citep{Miao2019_ApJ}, coronal holes \citep{Bagashvili2018_ApJL}, and the quiet Sun
\citep{Panesar2017_ApJ}.
Recurrent jets provide a natural setting for studying how the magnetic configuration,
energy, and helicity evolve between successive eruptions.
Previous studies have suggested that photospheric magnetic flux cancellation and emergence can
contribute to recurrent jet formation (\citealp{Chen2015_ApJ};
\citealp{Liujiajia_2016ApJ}; \citealp{Panesar2016_ApJL}; \citealp{Miao2019_ApJ}).
However, several fundamental questions remain open.
For example, how does the coronal magnetic system rebuild itself after one eruption and become
capable of producing the next?
Which properties of the magnetic configuration persist from one eruption to the next, and how are
magnetic energy and helicity redistributed between successive jets?

The initiation and evolution of solar coronal jets involve various magnetic structures and
topologies, including twisted magnetic flux ropes \citep{Zhu2017_ApJL}, sheared arcades
\citep{Wyper2017_Nature}, null points \citep{Pariat2015_AA}, hyperbolic flux tubes (HFTs;
\citealp{Sun2023_ApJ}), and current sheets \citep{Pariat2015_AA}.
Several complementary diagnostics have been developed to characterize 3D magnetic structures and
their evolution.
For example, twist and writhe \citep{Berger2006_JPhA} quantify the winding of magnetic field lines
around the flux-rope axis and the coiling of the axis itself, respectively.
These quantities are useful for studying flux-rope stability and the evolution of magnetic geometry
during eruptions \citep{Torok2010_AA}.
The squashing factor $Q$ quantifies the distortion of magnetic-field-line mapping, with high-$Q$
regions corresponding to quasi-separatrix layers (QSLs), which are characterized by strong gradients
in field-line connectivity \citep{Titov2002_JGR} and can be favorable sites for current-sheet
formation and magnetic reconnection \citep{Guo2013_AA}.

Complementing these diagnostics, magnetic helicity provides a quantitative measure of the
topological complexity of magnetic fields \citep{Berger1984_JFM}
and can help characterize their eruptive potential \citep{Nindos2004_ApJ}.
Using flux-emergence simulations, \citet{Pariat2017_AA} identified the ratio of
current-carrying to total relative magnetic helicity, $|H_J|/|H_V|$, as a
promising diagnostic of eruptivity.
Moreover, in four closely related line-tied eruption simulations with different photospheric drivers,
\citet{Linan_2020A&A} found torus-instability onset near a model-specific value of the absolute
helicity ratio $|H_J|/|H_V|=0.29\pm0.01$, although the systems reached this value through different
helicity-evolution pathways.
On the observational side, \citet{Zhou2025_ApJ} found that $H_J/H_V$ increased before and decreased
during three recurrent jets in one active region, approximately in phase with the EUV emission.
\citet{Bian2026_ApJ} found that $H_J/H_V$ clustered around $0.38\pm0.04$ at eruption onset in twelve
reconnection-triggered simulations with different magnetic configurations and photospheric drivers.
However, whether such onset values remain robust to the choice of integration volume requires
further examination.

In this paper, we perform a 3D thermodynamic MHD simulation of recurrent coronal jets driven by
magnetic flux cancellation induced by converging flows at the bottom boundary.
The simulation naturally produces two successive jet eruptions, allowing a direct comparison of
their magnetic topology, energy evolution, and helicity properties.

\section{Numerical Model} \label{sec:2}

\begin{figure*}[t]
\centering
\includegraphics[width=\textwidth]{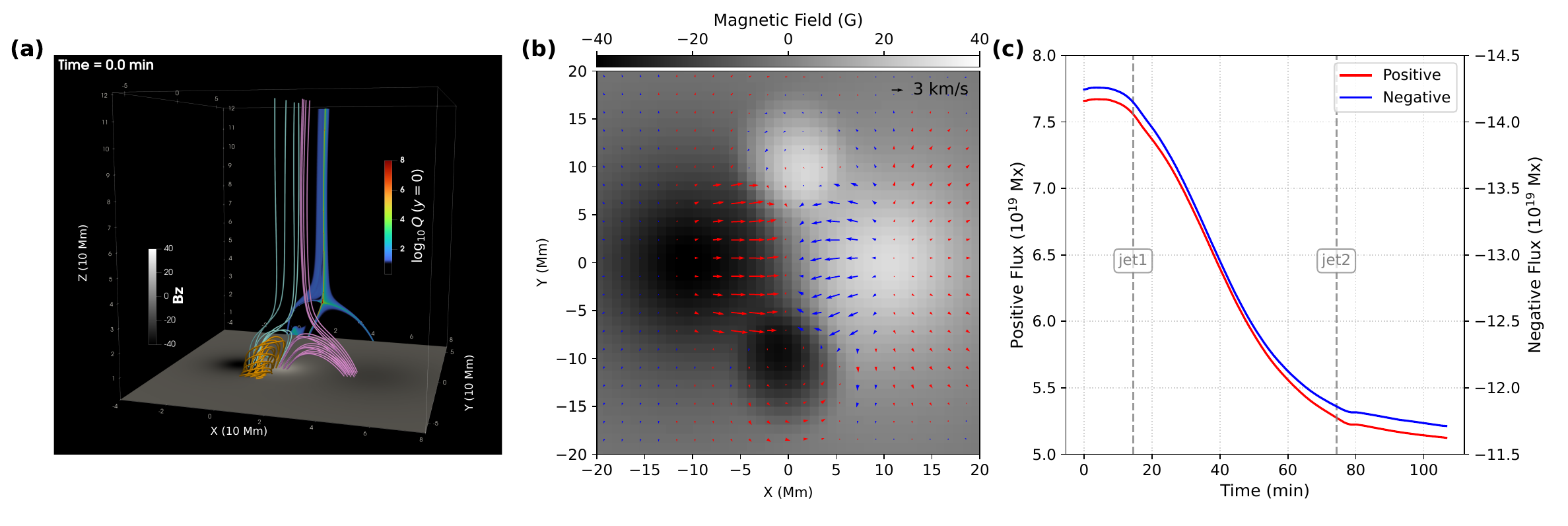}
\caption{
(a) Initial magnetic field configuration.
The flux rope is shown in orange, the overlying confining field in blue, and the surrounding field
in pink.
The squashing factor $Q$ is computed on the $y=0$ plane and displayed on a slice shifted to
$y=60$~Mm for visualization purposes.
(b) Spatial distribution of the imposed converging velocity at the bottom boundary.
(c) Temporal evolution of the magnetic flux within the selected region shown in panel (b).
The vertical dashed lines indicate the onset times of the first and second jet eruptions.
}
\label{fig:1}
\end{figure*}

We adopt a 3D thermodynamic MHD model to simulate recurrent jets using the parallelized Adaptive
Mesh Refinement Versatile Advection Code (\href{https://amrvac.org/}{MPI-AMRVAC};
\citealp{Keppens2012_JCP}; \citealp{Xia2018_ApJS}; \citealp{Keppens2023_AA}).
The computational domain is a Cartesian box spanning $x\in[-40,80]~\mathrm{Mm}$,
$y\in[-60,60]~\mathrm{Mm}$, and $z\in[0,240]~\mathrm{Mm}$.
The upper boundary is placed at $z=240~\mathrm{Mm}$ to allow the jets to propagate outward while
minimizing its influence on their evolution.
All subsequent visualizations and diagnostic calculations are restricted to
$z\in[0,120]~\mathrm{Mm}$. We use a base grid of $32\times32\times64$ cells
and apply four AMR levels, with a finest spatial resolution of
$468.75~\mathrm{km}$ in each direction.

The numerical fluxes are computed with the Harten--Lax--van Leer (HLL) approximate Riemann solver,
combined with the Cada3 limiter.
Time integration uses a three-step Runge--Kutta scheme, and the solenoidal constraint on the
magnetic field is maintained using constrained transport (CT).

The energy equation includes thermal conduction, optically thin radiative losses, and empirical
coronal heating.
The governing equations are as follows:

\begin{eqnarray}
\frac{\partial \rho}{\partial t} +\nabla \cdot(\rho \boldsymbol{v})=0\\
\kern-10pt\frac{\partial(\rho\boldsymbol{v})}{\partial t} +
    \nabla \cdot [\rho\boldsymbol{v}\boldsymbol{v} + (p+\frac{\boldsymbol{B}^{2}}{2\mu_{0}})\boldsymbol{I} - \frac{\boldsymbol{BB}}{\mu_0}]\kern10pt
    &\kern-10pt= \rho \boldsymbol{g}\kern10pt\\
\frac{\partial \boldsymbol{B}}{\partial t} + \nabla \cdot(\boldsymbol{vB-Bv})=0\\
\frac{\partial e_{\rm int}}{\partial t}+\nabla \cdot(e_{\rm int} \boldsymbol{v}) =- p \nabla \cdot
    \boldsymbol{v} + \eta J^{2} +\rho \boldsymbol{g \cdot v}+\\H_{0}e^{-z/\lambda}-n_{\rm e}n_{\mathrm{H}}\Lambda(T)
\notag
 + \nabla \cdot(\boldsymbol{\kappa} \cdot \nabla T)
\end{eqnarray}
where $p$ denotes the thermal pressure;
$\boldsymbol{g}=-g_{\odot}r_{\odot}^2/(r_{\odot}+z)^2\boldsymbol{e_{z}}$ is the gravitational
acceleration, with $g_{\odot}= \rm 274\ m\ s^{-2}$ being its value on the solar surface and
$r_{\odot}$ the solar radius; $e_{\rm int} =p/(\gamma -1)$ denotes the internal energy density;
$\boldsymbol{\kappa}=\kappa_{\parallel}\boldsymbol{\hat{b}\hat{b}}$ describes thermal conduction
along magnetic fields, where $\kappa_{\parallel}(T)=10^{-6}T^{5/2}~\mathrm{erg\,cm^{-1}\,s^{-1}\,K^{-1}}$
is the Spitzer thermal conductivity;
$n_{\rm e}n_{\rm {H}}\Lambda(T)$ represents the optically thin radiative loss term;
and $H_0e^{-z/\lambda}$ is an empirical heating term introduced to maintain the high coronal
temperature, with $H_{0}=10^{-4} \rm \ erg \ cm^{-3}\ s^{-1}$ and $\lambda=60\ \rm Mm$; $\eta J^{2}$
represents the Joule heating, where $\eta=1.0\times 10^{-4}$ in nondimensional units.

The initial magnetic field combines a modified Titov–Démoulin (TDm) flux-rope model
\citep{Titov2014} with a set of subphotospheric magnetic charges, which are
$q_1=-0.57~\mathrm{T\,Mm^2}$ at $(-10,0,-10)~\mathrm{Mm}$,
$q_2=0.47~\mathrm{T\,Mm^2}$ at $(10,0,-10)~\mathrm{Mm}$, and
$q_3=-1.2~\mathrm{T\,Mm^2}$ at $(50,0,-30)~\mathrm{Mm}$.
This configuration produces a magnetic topology consisting of a stable flux rope, a coronal null
point, and surrounding open magnetic field lines (Figure~\ref{fig:1}(a)).
We relax the initial magnetic field using the magneto-frictional model \citep{Guo2016_ApJa,
Guo2016_ApJb} to obtain a sufficiently force-free configuration.
Unlike the model of \citet{Pariat_2010ApJ}, our configuration favors a blowout jet driven by a
minifilament eruption.

We initialize the atmosphere in hydrostatic equilibrium, with a piecewise temperature profile
spanning the chromosphere, transition region, and corona.
Details of the atmospheric setup are given in \citet{Guo2024_ApJ}.

At the four side boundaries and the top boundary, the velocity in the ghost cells is set
by constant extrapolation from the interior, while the magnetic field is set using second-order
zero-gradient extrapolation.
At the top boundary, the density and gas pressure are extrapolated assuming hydrostatic equilibrium.
At the four side boundaries, both quantities are set by constant extrapolation.

To drive the system toward eruption, converging flows are imposed at the bottom boundary ($z=0$), as
shown in Figure~\ref{fig:1}(b).
We adopt a converging speed of approximately $3~\mathrm{km\,s^{-1}}$, slightly higher than typical
observed values, to keep the computation affordable while limiting the influence of numerical diffusion.
These flows transport opposite-polarity magnetic flux toward the polarity inversion line, leading to
magnetic flux cancellation and the gradual buildup of electric current.

\section{Results} \label{sec:results}
Converging motions imposed at the lower boundary drive sustained magnetic flux cancellation, with
the unsigned flux of both polarities decreasing nearly synchronously
(Figure~\ref{fig:1}(c)).
Starting from the initial TDm flux-rope configuration, the system produces two successive jets.
We define the onset of each jet as the time when the kinetic energy in the domain
begins to increase rapidly.
The onset times are $t=14.5$ and $74.3$~min for the first and second jets, respectively.
These times are marked by vertical dashed lines in Figure~\ref{fig:1}(c) and subsequent relevant
figures.
The first jet occurs during the early phase of flux cancellation, and the second follows
approximately one hour later.
Following the second eruption, the magnetic flux curves level off near $t \approx 80$ minutes,
indicating that flux cancellation has largely ceased.
No further jet eruptions occur during the remaining simulation interval.

\subsection{Overview of the Recurrent Jet Evolution}

\begin{figure*}[t]
\centering
\includegraphics[width=\textwidth]{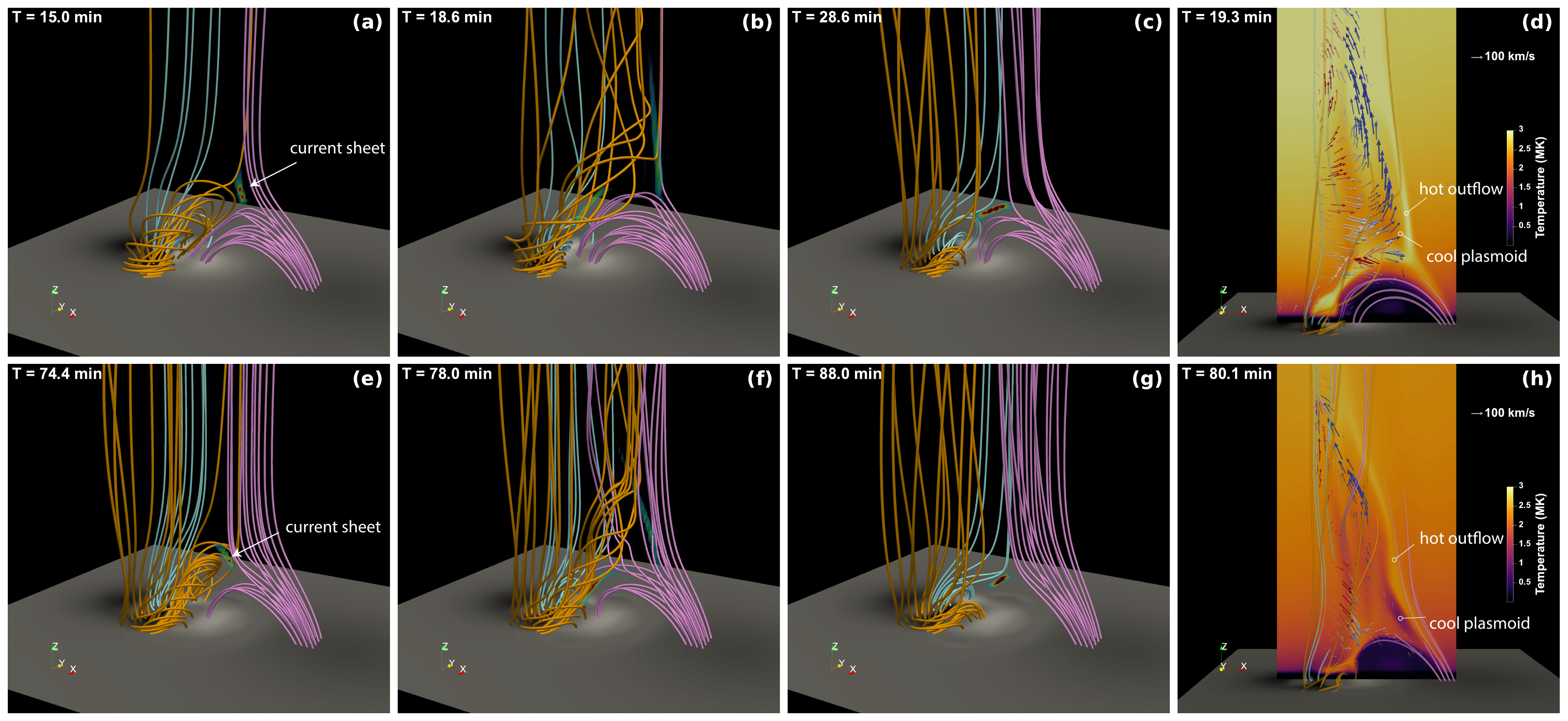}
\caption{
Evolution of the first jet (top row) and the second jet (bottom row).
Panels (a)--(c) and (e)--(g) show the magnetic field configuration at different times.
The flux rope is shown in orange, the overlying confining field in blue, and the surrounding
magnetic field in pink.
The rainbow slice represents the distribution of $J/B$ on $y=0$, used to trace the current sheet.
Panels (d) and (h) show the temperature distribution overlaid with velocity vectors on the $y=0$
slice at the eruption peaks.
}
\label{fig:jets}
\end{figure*}

Although both events are blowout jets, their magnetic configurations differ noticeably.
Compared with the flux rope before the first eruption, the flux rope before the second eruption
is more strongly twisted and has a smaller radius.
We also display the distribution of $J/B$ in the $y=0$ plane in Figure~\ref{fig:jets} as a proxy
for current sheets, highlighting possible sites of magnetic reconnection.

At a simulation time of $t=15$ minutes, shortly after the onset of the first jet, a current sheet
is visible (Figure~\ref{fig:jets}(a)).
Figure~\ref{fig:jets}(b) shows the blowout phase of the first jet.
During this stage, reconnection between the expanding flux rope and the surrounding field (pink
field lines) becomes more efficient, leading to the formation of an extended current sheet.
As a result, part of the flux rope reconnects into open field lines, facilitating the release of
magnetic twist, while part of the open field reconnects into closed arcade structures.

After the first eruption, a residual current sheet persists, enabling continued reconnection
(Figure~\ref{fig:jets}(c)).
This ongoing reconnection gradually reorganizes the magnetic field into the configuration preceding the second eruption.

The second eruption follows a similar sequence of flux-rope rise, reconnection, and untwisting.
However, its final magnetic configuration differs from that following the first eruption.
At $t=88.0$ minutes (Figure~\ref{fig:jets}(g)), more of the field lines traced in orange and
pink are open than at $t=28.6$ minutes (Figure~\ref{fig:jets}(c)), suggesting less magnetic flux
remaining in the residual flux rope.

The current sheets are located at different heights near the onset of the two eruptions.
Estimates from the $J/B$ distributions on the $y=0$ plane indicate heights of approximately
$20~\mathrm{Mm}$ for the first eruption and $15~\mathrm{Mm}$ for the second.

The second jet has a lower plasma temperature and speed than the first, as illustrated by the
temperature slices and velocity arrows in the $y=0$ plane at the eruption peaks
(Figure~\ref{fig:jets}(d) and (h)).
Both jets exhibit a characteristic multi-thermal structure, including hot reconnection outflows on
one side and a relatively cool plasmoid associated with the partially heated flux rope.
Arrow length represents the in-plane speed, whereas red and blue indicate positive and negative
values of $v_y$, respectively.
The spatially separated oppositely directed velocity components are consistent with rotational
motion, qualitatively resembling the velocity pattern inferred from observations of rotating jets
(\citealp{Liujiajia_2018ApJ}; \citealp{Huang2026_ApJ}).
The lower temperature and speed of the second jet may reflect a smaller magnetic energy release
during that eruption.

\begin{figure}
\centering
\includegraphics[width=\columnwidth]{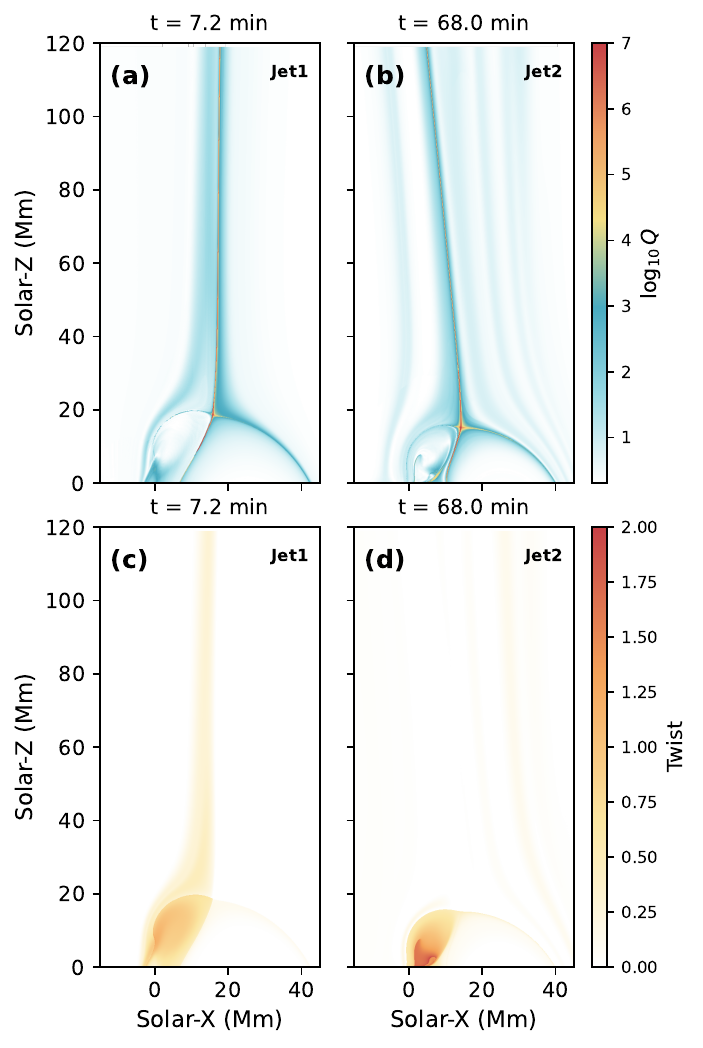}
\caption{
Distributions of the squashing factor $Q$ in panels (a) and (b) and magnetic twist $T_w$ in panels
(c) and (d), calculated on the $y=0$ slice prior to the two eruptions.
}
\label{fig:q_twist}
\end{figure}

We identify QSLs using the squashing factor $Q$ \citep{Titov2002_JGR}.
The magnetic twist is quantified by the twist number $T_w$, defined by \cite{Berger2006_JPhA} as
\begin{equation}
T_w = \int_L \frac{(\nabla \times \mathbf{B}) \cdot \mathbf{B}}{4\pi B^2}\, dl,
\end{equation}
where $L$ is the integration path along a magnetic field line.
The twist number $T_w$ measures the number of turns that magnetic field lines wind about
infinitesimally close field lines.
Figure~\ref{fig:q_twist} shows the distributions of the QSLs and magnetic twist on the $y=0$ slice, computed using the \texttt{FastQSL} code (\citealp{Zhang2022_ApJ}).
The QSL maps show that the outer spine-like structure is more inclined before Jet~2 than before Jet~1, suggesting that the first eruption has modified the geometry of the surrounding magnetic field.
In addition, both the flux rope and the enclosing dome-like QSL structure are more compact before
Jet~2.
This difference in flux-rope morphology is also evident during the early stages of the eruptions
(Figure~\ref{fig:jets}(a) and (e)).
The more compact configuration before Jet~2 may be related to the lower current-sheet height
during the second eruption.

The magnetic-twist distributions also differ clearly between the two jets.
Before the first eruption, the flux rope has a twist of approximately unity over most of its cross
section.
But before the second eruption, the lower portion of the rope reaches twist values of up to $\sim2$,
while its upper portion also retains relatively high twist.
In addition, the enhanced twist in the surrounding field shifts from the left side of the system
before Jet~1 to the right side before Jet~2, indicating a reorganization of the overall magnetic
topology.

\subsection{Energy and Helicity Evolution}
The magnetic energy and free magnetic energy in the reference analysis volume are calculated as
\begin{equation}
E = \int_V \frac{B^2}{8\pi} \, dV,
\end{equation}
\begin{equation}
E_{\mathrm{free}} = E - E_{\mathrm{pot}},
\end{equation}
where $E_{\mathrm{pot}}$ is the magnetic energy of the corresponding potential field with the same
normal component at the boundary.

\begin{figure}[!t]
\centering
\includegraphics[width=\columnwidth]{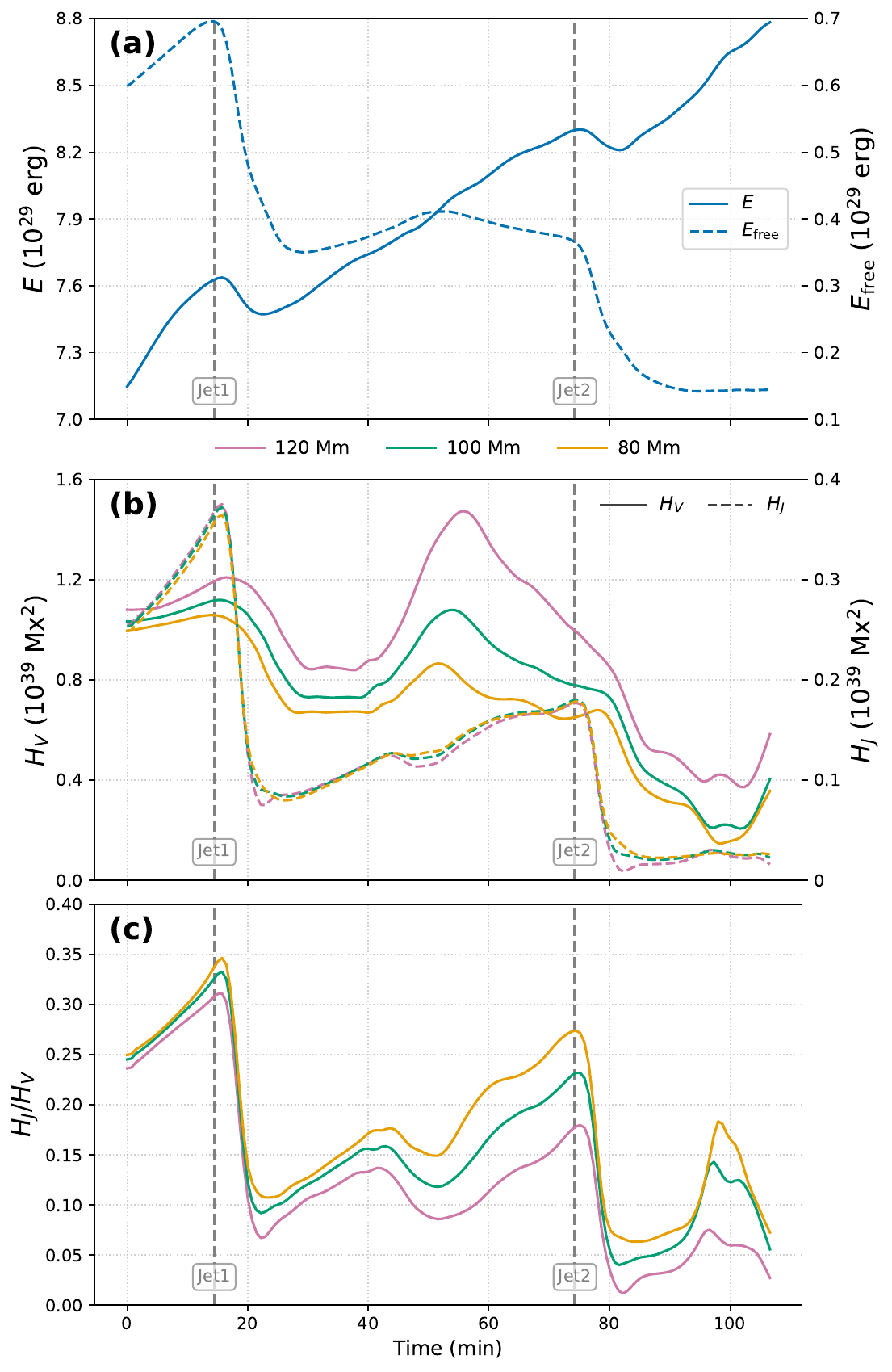}
\caption{
Temporal evolution of magnetic energy and helicity.
(a) Total magnetic energy $E$ and free magnetic energy $E_{\mathrm{free}}$ in the reference analysis
volume.
(b) Relative magnetic helicity $H_V$ (solid lines) and current-carrying helicity $H_J$ (dashed
lines).
(c) Helicity ratio $H_J/H_V$.
In panels (b) and (c), the pink, green, and orange curves correspond to boxes with side lengths of
120, 100, and 80~Mm, respectively.
The 120~Mm box is the reference analysis volume spanning $z\in[0,120]~\mathrm{Mm}$, while the 100
and 80~Mm boxes are reduced volumes encompassing the core magnetic structure.
The gray vertical dashed lines indicate the onset times of the two jet eruptions.}
\label{fig:E_helicity}
\end{figure}

The total and free magnetic energies exhibit different evolutionary trends
(Figure~\ref{fig:E_helicity}(a)).
The total magnetic energy increases overall under the continuous bottom-boundary driving but
decreases slightly during each jet eruption.

By contrast, the free magnetic energy shows a distinct evolution pattern.
First, the free magnetic energy decreases more markedly than the total magnetic energy during
both eruptions. Second, unlike the total magnetic energy, the free magnetic energy does not increase continuously
between the eruptions, despite the marked changes in magnetic topology shown in
Figure~\ref{fig:jets}.
These results suggest that free magnetic energy responds clearly to eruptions but,
on its own, provides limited information about the detailed topological evolution and the onset
conditions of subsequent eruptions.

After the second eruption, the free magnetic energy remains low and nearly constant.
Together with the evolution of the magnetic flux and topology, this suggests that flux cancellation
may no longer be sufficient to form a new flux rope and accumulate enough free magnetic energy to
trigger a third eruption.

The gauge-invariant relative magnetic helicity was introduced for magnetically open volumes by
\citet{Berger1984_JFM} and is written here in the Finn--Antonsen form \citep{FinnAntonsen1985_CPCF}.
The current-carrying helicity follows \citet{Berger1999_PPCF}.
Both quantities are commonly used to characterize the complexity of magnetic field topology.
They are defined as
\begin{equation}
H_V = \int_V (\mathbf{A} + \mathbf{A}_{\mathrm{p}})\cdot(\mathbf{B} - \mathbf{B}_{\mathrm{p}})\, dV,
\end{equation}
\begin{equation}
H_J = \int_V (\mathbf{A} - \mathbf{A}_{\mathrm{p}})\cdot(\mathbf{B} - \mathbf{B}_{\mathrm{p}})\, dV,
\end{equation}

\noindent where $\mathbf{B}$ is the magnetic field of the volume,
$\mathbf{B}_{\mathrm{p}}$ is the corresponding potential field with the same normal component on the
boundary, and $\mathbf{A}$ and $\mathbf{A}_{\mathrm{p}}$ are corresponding vector potentials
satisfying $\mathbf{B} = \nabla \times \mathbf{A}$ and
$\mathbf{B}_{\mathrm{p}} = \nabla \times \mathbf{A}_{\mathrm{p}}$.
$H_J$ represents the helicity associated with the non-potential (current-carrying) part of the
magnetic field.

Previous observational and numerical studies commonly calculate magnetic helicity within a chosen volume.
This choice is not unique and may affect the resulting helicity diagnostics.
This important but often overlooked issue means that helicity-based onset thresholds
inferred from a single volume may not be robust.
We therefore compare the helicity evolution across different integration volumes.

We use three nested volumes of nominal size 120, 100, and 80~Mm, sharing the same horizontal center
and lower boundary and enclosing the core magnetic structure.
Each volume is extracted from the same magnetic snapshots without changing the grid spacing.
The reference potential field is recomputed from the normal field on each volume's six faces,
and the vector potentials and helicities are then evaluated independently within that volume.

We get some interesting results.

First, $H_V$ and $H_J$ respond differently to the integration volume.
Across the tested volumes, which all enclose the core magnetic structure, $H_V$ generally
increases with volume, whereas $H_J$ varies much less (Figure~\ref{fig:E_helicity}(b)).
This suggests that $H_J$ is dominated by the current-carrying core flux rope, while $H_V$ also
depends on the surrounding magnetic field.

Second, $H_J$ provides information about magnetic evolution beyond that captured by
$E_{\mathrm{free}}$.
It decreases clearly during each eruption but gradually
recovers between the two jets.
This recovery is consistent with the rebuilding of the flux rope before the second eruption,
but the changes in $E_{\mathrm{free}}$ do not reflect this rebuilding.

Third, $H_J/H_V$ changes markedly around both eruptions (Figure~\ref{fig:E_helicity}(c)),
supporting its close association with eruptive activity
\citep{Gupta2021_AA,Duan2023_ApJ,Pariat2023_AA}.
However, our results do not establish a common pre-eruptive threshold, even for recurrent jets from
the same source region with similar eruptive structures.
Within the same 120~Mm volume, the ratio reaches markedly different peaks near the onset of
the two eruptions, with values of $\sim0.31$ and $\sim0.18$
(pink curve in Figure~\ref{fig:E_helicity}(c)).
These peaks are not strictly pre-eruptive measurements.
Moreover, Figure~\ref{fig:E_helicity}(c) shows that larger integration volumes generally yield
smaller $H_J/H_V$, reflecting the volume dependence of $H_V$.

Together, $E_{\mathrm{free}}$, $H_J$, and $H_J/H_V$ provide useful, complementary diagnostics of
eruptive dynamics, but each has limitations when used alone to diagnose eruption onset.

\section{Discussion}\label{sec:discussion}

The usefulness of a helicity-based quantity for tracking eruptive activity should be distinguished
from its ability to provide a predictive onset threshold.
The narrow range of onset values reported by \citet{Bian2026_ApJ} indicates consistency in this
ratio across their simulation ensemble.
By contrast, the peak values we obtain near eruption onset differ from their reported values,
suggesting that the characteristic ratio may be case-dependent.
More importantly, we find that even the same magnetic state yields different helicity ratios
when evaluated over different integration volumes.
Applying such numerical thresholds therefore requires assessing their applicability across
different cases and accounting for the choice of integration domain and its effects.

\begin{figure}[htbp]
\centering
\includegraphics[width=\columnwidth]{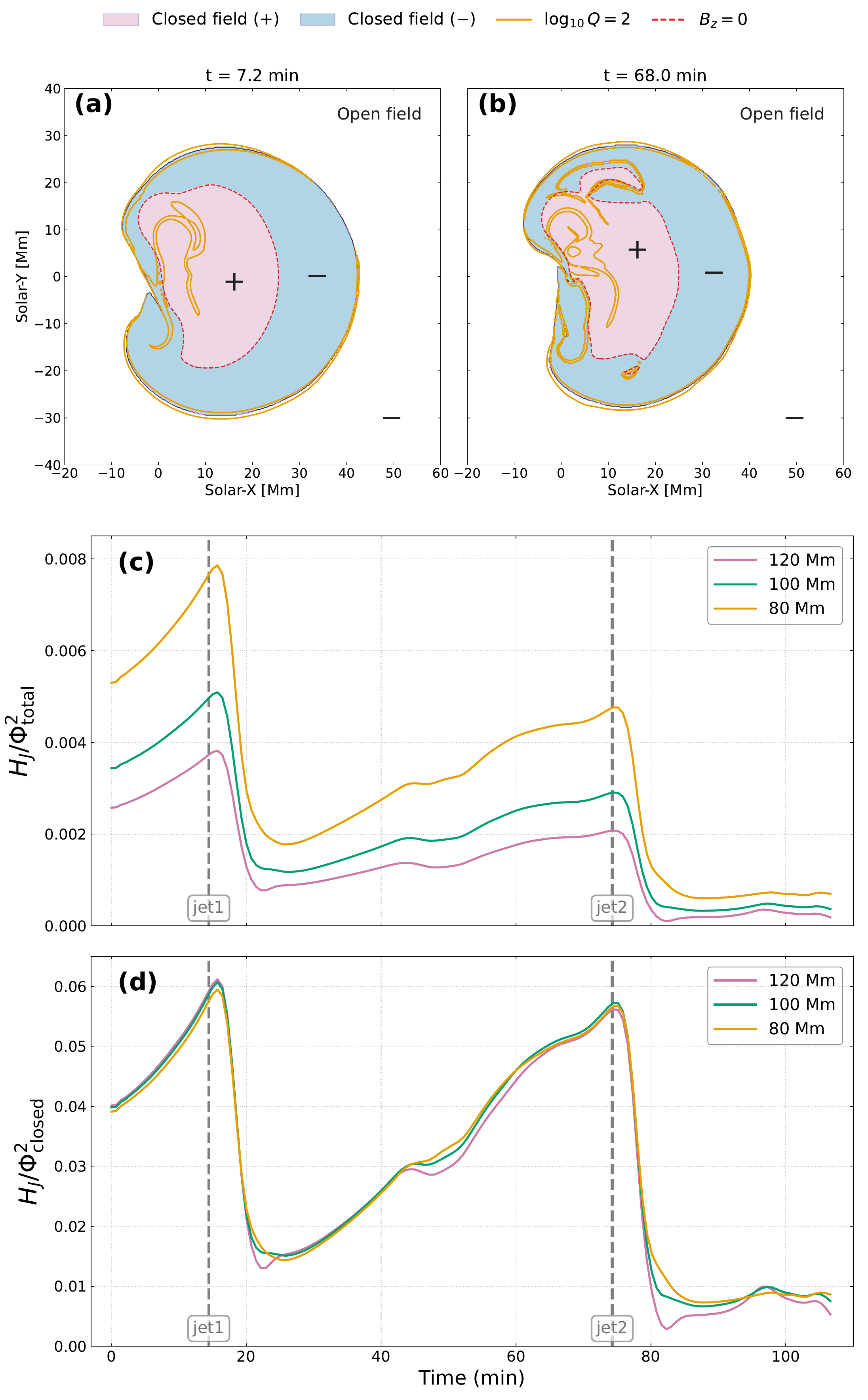}
\caption{
Bottom-boundary magnetic connectivity and normalized current-carrying helicity.
(a) and (b) Closed-field footpoint maps at $t=7.2$~min and $t=68.0$~min, respectively.
Field lines are classified as closed when both ends return to the bottom boundary.
Pink and blue shading distinguish their positive ($B_z>0$) and negative ($B_z<0$) footpoints,
respectively; the surrounding open-field region is left white.
The $+$ and $-$ symbols indicate the sign of the bottom-boundary $B_z$.
Orange solid and red dashed contours mark $\log_{10}Q=2$ and $B_z=0$, respectively.
(c) and (d) Temporal evolution of $H_J/\Phi_{\mathrm{total}}^2$
and $H_J/\Phi_{\mathrm{closed}}^2$, respectively,
for the 120, 100, and 80~Mm integration
volumes, shown in pink, green, and orange, respectively.
In panel (c), $\Phi_{\mathrm{total}}$ is half of the total unsigned magnetic flux
through the bottom boundary of each integration volume.
All three curves in panel (d) use the same $\Phi_{\mathrm{closed}}$, the single-polarity magnetic flux of the
closed-field domain.
The gray vertical dashed lines indicate the onset times of the two jet eruptions.}
\label{fig:Hj_phi}
\end{figure}

The volume dependence of the global ratio can be understood through the decomposition
$H_V=H_J+H_{PJ}$, where $H_{PJ}$ describes the cross contribution involving the reference potential
field and the current-carrying field \citep{Linan_2020A&A}.
Changing the integration volume changes both the domain of integration and the reference potential
field determined by its boundary conditions.
Thus, $H_J/H_V$ reflects not only the current-carrying structure but also its relationship with the
surrounding magnetic field.

An alternative is to normalize $H_J$ by the magnetic flux associated with the eruptive source region.
Previous studies have investigated flux-normalized helicity quantities using half of the total
unsigned photospheric magnetic flux \citep{Duan2023_ApJ}.
We also test this normalization, as shown in Figure~\ref{fig:Hj_phi}(c),
but the near-onset peaks differ between the two eruptions and remain sensitive to the integration volume.
In fact, $\Phi_{\mathrm{total}}$ varies by only about 10\%, so the shape of
$H_J/\Phi_{\mathrm{total}}^2$ remains primarily determined by $H_J$.
We speculate that this is because the flux may include contributions from outside the eruptive core.
Here, we instead consider $H_J/\Phi_{\mathrm{closed}}^2$, where $\Phi_{\mathrm{closed}}$ is the
single-polarity magnetic flux of the closed-field domain identified from magnetic connectivity
(Figures~\ref{fig:Hj_phi}(a) and~\ref{fig:Hj_phi}(b)).
This quantity reaches comparable peaks near the onset of both eruptions, with values of
approximately $0.06$ and $0.055$,
which are relatively insensitive to the choice among the tested integration volumes, despite the
reduction in closed magnetic flux and the reconfiguration of the field between them
(Figure~\ref{fig:Hj_phi}(d)).
This agreement suggests that the flux content of the source region provides a relevant scale for
comparing the helicity associated with successive eruptions.

The physical relevance of this normalization depends on how the closed-field domain is selected.
In the present configuration, the outer QSL footprint closely follows the boundary of the
closed-field region containing the principal non-potential structures
(Figures~\ref{fig:Hj_phi}(a) and~\ref{fig:Hj_phi}(b)).
This correspondence provides a topological basis for the adopted flux selection.
In more complex magnetic configurations, however, several closed-field systems may coexist, and
their combined flux need not represent the eruptive source.
The QSLs surrounding the eruptive core may help identify an appropriate domain.
However, calculating the QSLs and defining the region can be complicated, and the choice of region
is somewhat subjective.

Although $H_J/\Phi_{\mathrm{closed}}^2$ reaches similar peaks near the onset of both eruptions,
its ability to predict eruption onset remains to be tested.
Further tests across different magnetic configurations and driving conditions are needed to assess
whether this diagnostic can reliably characterize eruptive conditions.
The computational methods also require further refinement for application to complex coronal
magnetic fields extrapolated from observations.

\section{Summary}\label{sec:summary}

We have performed a 3D thermodynamic MHD simulation in which converging motions at the bottom
boundary drive magnetic flux cancellation and produce two successive blowout jets.

Both $H_J$ and $E_{\mathrm{free}}$ decrease during the eruptions.
Between the two jets, however, $H_J$ gradually recovers, whereas $E_{\mathrm{free}}$ shows no clear
net buildup.
The recovery of $H_J$ is consistent with flux-rope rebuilding, highlighting the complementary
information provided by energy and helicity.

Across the tested volumes enclosing the core magnetic structure, $H_J$ varies less than $H_V$, and
$H_J/H_V$ depends on volume selection.
Even within the same 120~Mm reference volume, the ratio reaches different peaks near the onset of
the two eruptions, with values of approximately $0.31$ and $0.18$.

By contrast, $H_J/\Phi_{\mathrm{closed}}^2$ reaches comparable peaks near the onset of both eruptions,
with values of approximately $0.06$ and $0.055$, which are relatively insensitive to the choice
among the tested integration volumes.
Here, $\Phi_{\mathrm{closed}}$ is the single-polarity magnetic flux of the closed-field domain.
These results suggest that local flux normalization may provide a useful diagnostic of eruptive
conditions, although its ability to predict eruption onset remains to be tested.

\begin{acknowledgments}
C.Z., Y.G., and J.H.G. were supported by the National Key R\&D Program of China (2022YFF0503004,
2021YFA1600504, and 2020YFC2201201), NSFC (12333009), and the Fundamental Research Funds for the
Central Universities (KG202506).
Z.L. is supported by JSPS Postdoctoral Fellowships.
J.L. is supported by the National Natural Science Foundation of China (12373056).
The numerical computation was conducted in the High Performance Computing Center (HPCC) at Nanjing
University.

\end{acknowledgments}

\bibliography{helicity}{}
\bibliographystyle{aasjournal}
\end{document}